\documentclass[11pt]{article}
\usepackage{amssymb}
\usepackage{graphicx}
\usepackage{natbib}
\usepackage{url}

\begin{document}

\begin{center}
{\large\bf
On the two main classes of Active Galactic Nuclei}
\medskip
\end{center}

{\bf\tt Paolo Padovani}
\medskip\medskip

{\bf Active Galactic Nuclei (AGN) are traditionally divided empirically
  into two main classes: ``radio-loud'' and ``radio-quiet'' sources. These
  labels, which are more than fifty years old, are obsolete, misleading,
  and wrong. I argue that AGN should be classified based on a fundamentally
  physical rather than just an observational difference, namely the presence (or
  lack) of strong relativistic jets, and that we should use the terms
  ``jetted'' and ``non-jetted'' AGN instead. }
  
\medskip

Everybody knows that AGN are powered by a supermassive black hole (SMBH).
And (almost) everybody knows that there are two main classes of AGN: the
``radio-loud'' (RL) and the ``radio-quiet'' (RQ) ones. These classifications go all the way
back to \cite{1965ApJ...141.1560S}, who realised, soon after the discovery
of the first quasar, 3C 273, a very strong radio source, that there were
many similar sources in the sky, which were however undetected by the radio
telescopes of the time. It was later understood that these quasars were
only ``radio-faint'', but the name stuck. Indeed, for the same optical
power the radio powers of RQ quasars are a few orders of magnitude smaller
than those of their RL counterparts. This is, in fact, how RQ quasars are
characterised: relatively low radio-to-optical flux density ratios ($R
\lesssim 10$) and radio powers \citep[$P_{\rm 1.4GHz} \lesssim 10^{24}$ W
  Hz$^{-1}$ locally: e.g.,][]{2016A&ARv..24...13P}.

We know now that RQ AGN are the norm, not the exception, as they make up
the large majority ($> 90\%$) of the AGN class
\citep[e.g.,][]{2011MNRAS.411.1547P}. We also know that, despite what the
odd labels might suggest, the differences between the two classes are not
restricted to the radio band, far from it. And they are not simply
taxonomic either, as the two classes represent {\it intrinsically}
different objects, with most RL AGN emitting a large fraction of their
energy non-thermally over the whole electromagnetic spectrum while the
multi-wavelength emission of RQ AGN is dominated by thermal emission,
directly or indirectly related to the accretion disk, which forms around
the SMBH.

The most striking difference is in the hard X-ray to $\gamma$-ray band:
while many (likely all: but see below) RL sources emit all the way up to
GeV ($2.4 \times 10^{23}$ Hz), and sometimes TeV ($2.4 \times 10^{26}$ Hz),
energies, nearby (RQ) bright Seyfert galaxies have a sharp cut-off at energies
$\lesssim 1$ MeV \citep[e.g.,][]{2014ApJ...782L..25M}. This cut-off has to apply to
the whole RQ AGN population in order not to violate the X-ray background
above this energies \citep{2005ExA....20...41C}. Moreover, no RQ AGN has
ever been detected in $\gamma$-rays \citep{2012ApJ...747..104A} with the
exception of NGC 1068 and NGC 4945, two Seyfert 2 galaxies in which the
$\gamma$-ray emission is thought to be related to their starburst component
\citep{2012ApJ...755..164A}. This means that, while RQ AGN are actually
      {\it not} radio-quiet, they are $\gamma$-ray-quiet. 

What are the differences between the two classes due to? One simple thing:
the presence (or absence) of a strong relativistic jet. The relative (and
absolute) strength of the radio emission in the two classes is just a
consequence of this fundamental {\it physical} difference. Hence the need
for new and better names: jetted and non-jetted AGN
\citep{2016A&ARv..24...13P}.

\begin{figure}
\centering
\vspace{-10.em}
\includegraphics[width=13cm]{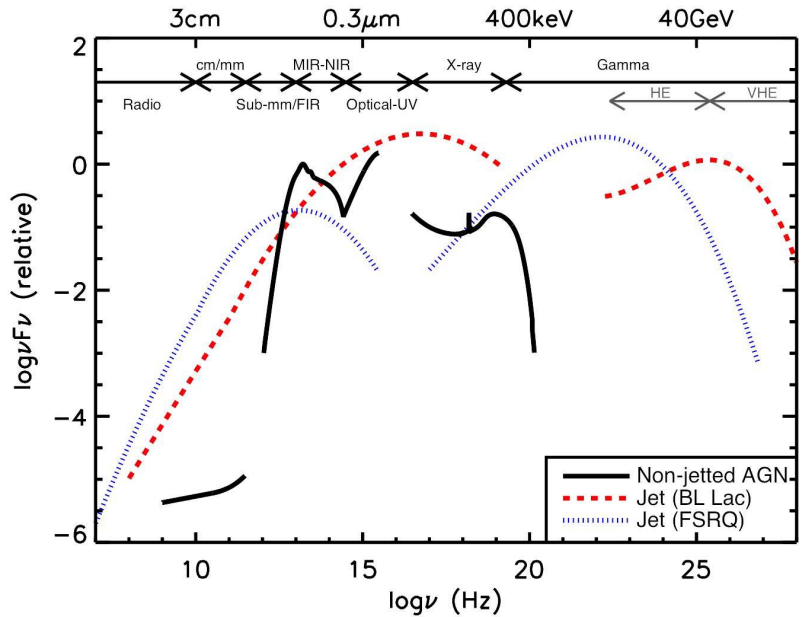}
\vspace{-3em}
\caption{A schematic representation of AGN SEDs. The black solid curve
  represents the typical SED of non-jetted AGN, while the red and blue
  lines refer to two jetted AGN, namely a BL Lac (based on the SED of Mrk
  421) and an FSRQ (based on the SED of 3C 454.3). Adapted from
  \cite{2014PhDT.......357H} and \cite{Padovani_2017}. Image
  credit: C. M. Harrison.}
\label{fig:AGN_SED}       
\end{figure}

This is illustrated in Fig. \ref{fig:AGN_SED}, which compares the spectral
energy distributions (SEDs) of typical non-jetted AGN with those of two
jetted ones, a BL Lac and a flat-spectrum radio quasar (FSRQ). Both of
these belong to the blazar class, which includes AGN with their jets
oriented at a very small angle ($\lesssim 15 - 20^{\circ}$) with respect to
the line of sight. Because of so-called unification models
\citep[e.g.,][]{1995PASP..107..803U}, the SEDs of blazars are
representative of {\it all} jetted sources as BL Lacs and FSRQs are
intrinsically the same sources as low-and high-excitation radio galaxies
(RGs) respectively, just seen at different angles \citep[see
  also][]{2014ARA&A..52..589H}. The SEDs of non-blazar, jetted AGN, are in
fact just shifted to lower frequencies by $\delta^{-1}$ (where
$\delta$ is the blazar Doppler factor; and to lower flux densities: Appendix B
of \citealt{1995PASP..107..803U}), and include also some extended,
isotropic emission in the radio band and the host galaxy in the optical
(typically swamped by the jet in blazars). As obvious from the Figure, the
SED of non-jetted AGN has a cutoff at much lower energies than those of
jetted AGN.

There are other reasons for dropping the old names. The classical
distinction between RL and RQ sources (based either on $R$ or radio power)
is valid {\it only} for broad-lined, unobscured AGN, i.e. quasars and
Seyfert 1s. Many RGs can have quite low $R$ or radio power values, at
levels typical of RQ AGN (\citealt{2011ApJ...740...20P,2016A&ARv..24...13P}
and in particular Fig. 4 of \citealt{2013MNRAS.436.3759B}). $R$ is useful
(and defined) for quasar samples, where it can be assumed that the optical
flux is related to the accretion disk and therefore the radio-to-optical
flux density ratio gives a measure of the jet/disk ratio. But it loses its
meaning as an indicator of jet strength if the optical band is dominated by
jet emission or by the host galaxy, as is the case in RGs. This has become
quite obvious only recently with the study of the source populations in
deep ($S_{\rm r} < 1$ mJy) radio fields, which have been detecting RL and
RQ sources in similar numbers and at similar flux densities and radio
powers \citep[e.g.,][]{2016A&ARv..24...13P}. Shall we
start calling RGs with $R < 10$ ``RQ AGN''? We should if we adhere to the
old scheme but this would not make any sense.

Moreover, in the case of Seyfert galaxies $R$ depends strongly on the
spatial resolution of the optical and radio observations.
\cite{2001ApJ...555..650H} have shown that by considering their nuclear
luminosities, most Seyfert 1 nuclei are RL, which further confuses the
issue. \cite{2003ApJ...583..145T} have introduced a new definition of the
radio loudness parameter by using the ratio between radio and $2 - 10$ keV
luminosity, $R_{\rm X}$. The idea was to avoid the extinction problems
affecting the optical band in obscured AGN, which would lead to
overestimated values of $R$. This might work if the $2 - 10$ keV luminosity
were produced by the same emission mechanism in all RL AGN but we know this
is not the case, as in some objects the X-ray luminosity can be dominated
by the jet itself \citep[e.g.,][]{2009MNRAS.396.1929H}.

How should the jetted and non-jetted AGN be defined? The RL/RQ division
might not have been very meaningful when applied outside the quasar class
but it had a simple (albeit of limited application) definition. I therefore
provide here some guidance, on how to distinguish between the two classes.

\begin{enumerate}

\item Direct evidence of a strong jet. This is obviously easy for bright
  radio sources, where one can identify strong, large, and resolved
  relativistic jets, also by following them up over time to detect
  superluminal motion, a clear sign of jet speeds getting close to the
  speed of light \citep[Appendix A of][]{1995PASP..107..803U}. However, in
  the deep radio fields discussed above most AGN have faint radio flux
  densities and are either unresolved or barely resolved.

\item $\gamma$-ray ($\gtrsim 1$ MeV) emission (Fig. \ref{fig:AGN_SED}), as
  only jetted AGN manage to reach these energies. But even the Large Area
  Telescope on board {\it Fermi} has limited sensitivity:
  the surface density of jetted AGN in the $\gamma$-ray band is about four
  orders of magnitude lower than in the radio band
  \citep{2016A&ARv..24...13P}, which shows that present-day $\gamma$-ray
  observatories only probe AGN with much higher bolometric fluxes compared
  to the current generation of radio telescopes.
  
\item Radio-excess off the far-infrared (FIR) -- radio correlation. The FIR
  and radio emission are strongly (and linearly) correlated in a variety of
  star-forming sources and non-jetted AGN \citep[e.g.,][and references
    therein]{2016A&ARv..24...13P}. Recent star formation is believed to be
  the driver of this correlation, at least in star-forming galaxies. Jetted
  AGN, because of their strong jet-related radio emission, display instead
  a ``radio excess'', which puts them off the correlation. Non-jetted AGN
  can have a radio core with flux density larger than their extended,
  likely star-formation-related component \citep[e.g.,][and references
    therein]{2016A&A...589L...3M}. Therefore, it somewhat matters where one
  draws the line (e.g., jetted AGN can be defined as being off the
  correlation by more than $2\sigma$, where $\sigma$ is the dispersion
  around the correlation). This radio-excess criterion, although indirect,
  is the simplest one to apply.

\end{enumerate}

I want to stress that jetted AGN are characterised by {\it strong,
  relativistic} jets. Non-jetted AGN can also have radio structures similar
to collimated outflows but these ``jets'' are small, weak, and slow
compared to those of jetted sources \citep[e.g.,][]{2004A&A...417..925M}.

Having understood the major difference between the two AGN classes, I think
we should now concentrate on the physics and the really big and
long-standing question: {\it Why are only a minority of AGN jetted}? We do
have some hints, as jetted AGN {\it appear} to be more clustered, undergo
mergers, reside in more massive, bulge-dominated galaxies (and perhaps spin
faster) than non-jetted AGN. All we have to do is take advantage of the
huge amount of radio data, which will be coming in the very near future
\citep[e.g.][and references therein]{2016A&ARv..24...13P} and answer it!

\medskip
{\em Paolo Padovani is at the European Southern Observatory, Karl-Schwarzschild-Str. 2,
D-85748 Garching bei M\"unchen, Germany\\
e-mail: ppadovan@eso.org
}

\small 

\bibliographystyle{spbasic_short}
\bibliography{NAT_comm} 

\medskip
{\bf Acknowledgments.} I thank Chiara Circosta, Chiara Ferrari, Evanthia
Hatziminaoglou, Andrea Merloni, Francesca Panessa, and Elisa Resconi for
helpful comments and Chris Harrison for producing Fig. \ref{fig:AGN_SED}.

\end{document}